\documentclass[9pt]{osa-supplemental-document}
\setboolean{shortarticle}{false}

\title{A Multi-Harmonic NIR-UV Dual-Comb Spectrometer}

\author[1]{Kristina F. Chang}
\author[1,2]{Daniel M. B. Lesko}
\author[1,3]{Carter Mashburn}
\author[1,3]{Peter Chang}
\author[1,4,†]{Eugene Tsao}
\author[1,4,†]{Alexander J. Lind}
\author[1,3,4,*]{Scott A. Diddams}

\affil[1]{Time and Frequency Division, National Institute of Standards and Technology, 325 Broadway, Boulder, CO 80305, USA.}
\affil[2]{Department of Chemistry, University of Colorado Boulder, Cristol Chemistry and Biochemistry, Boulder, CO 80309, USA.}
\affil[3]{Department of Physics, University of Colorado Boulder, Libby Dr, Boulder, CO 80302, USA.}
\affil[4]{Department of Electrical, Computer, and Energy Engineering, University of Colorado Boulder, 425 UCB 1B55, Boulder, CO 80309, USA.}

\affil[*]{Corresponding author: scott.diddams@colorado.edu}

\begin{abstract} 
Dual-comb spectroscopy in the ultraviolet (UV) and visible would enable broad bandwidth electronic spectroscopy with unprecedented frequency resolution. However, there are significant challenges in generation, detection and processing of dual-comb data that have restricted its progress in this spectral region. In this work, we leverage robust 1550 nm few-cycle pulses to generate frequency combs in the UV-visible. We couple this source to a wavelength multiplexed dual-comb spectrometer and simultaneously retrieve 100 MHz comb-mode-resolved spectra over three distinct harmonics spanning 380-800~nm. The experiments highlight the path to continuous dual-comb coverage spanning 200-750~nm, offering extensive access to electronic transitions in atoms, molecules, and solids.

\end{abstract}

\begin{document}
\maketitle

\begin{figure*}[ht!]
\centering
\includegraphics[width=\textwidth]{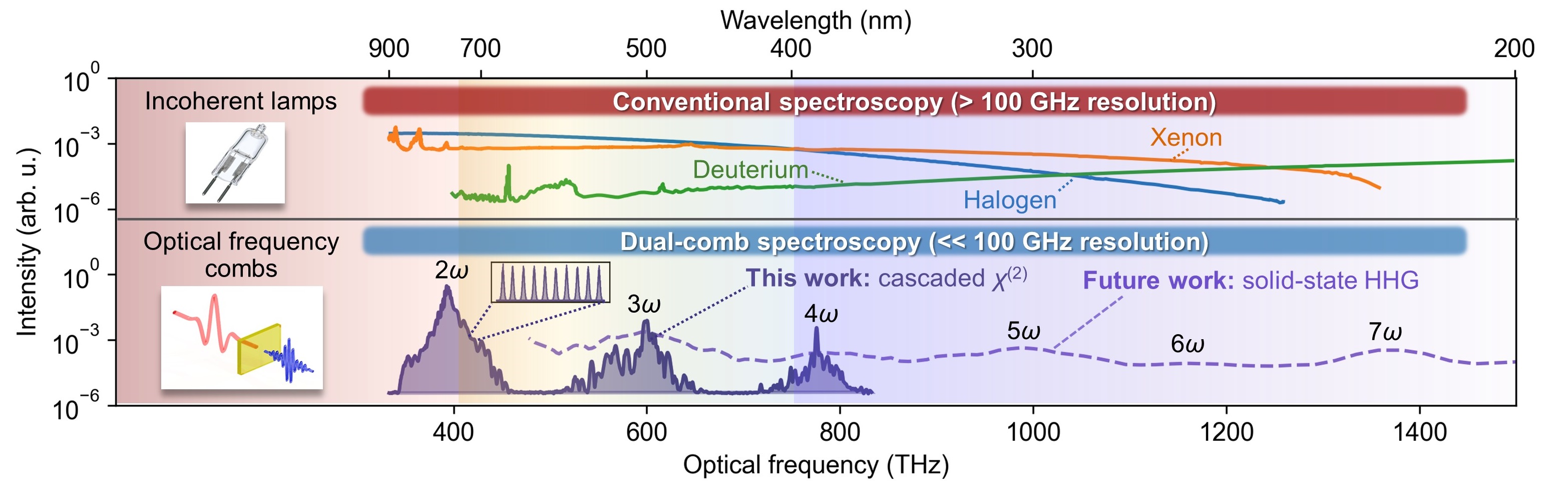}
\caption{Light sources for spectroscopy in the UV (750-1500~THz), visible (400-750~THz), and NIR regions (<400~THz), shaded in red, orange-blue, and purple, respectively. Lamp spectra (blue, orange, green curves) are adapted from Ref. \cite{Das:2021}. A NIR-UV frequency comb (solid purple curve) is generated through few-cycle pulse harmonic generation in a periodically-poled lithium niobate crystal. The few-cycle setup also extends to solid-state high-harmonic generation (HHG) in zinc oxide, providing complete UV-visible coverage (dashed purple curve). The relative scaling of the comb spectra reflects their intensities.}
\label{fig:1}
\end{figure*}

Ultraviolet (UV) and visible spectroscopy of electronic transitions is an essential tool in chemistry, biology, and materials science \cite{Galtier:2020,Das:2021}. Grating-based spectrometers are typically used to measure absorption spectra in the UV-visible from 200-750~nm (400-1500~THz). Broadband measurements are crucial for fully characterizing electronic spectra, which often span hundreds of THz. Conventional spectrometers commonly employ multiple incoherent lamps to cover the UV-visible range (Fig. \ref{fig:1}). However, these instruments are generally limited to >100 GHz resolution, hindering accurate spectral analyses required for applications from fundamental spectroscopy to remote sensing \cite{Galtier:2020, Schuster:2021}.

Dual-comb spectroscopy (DCS) is an emerging technique with intrinsic advantages in resolution and precision over existing spectroscopy modalities \cite{Coddington:2016,Weichman:2019,Picque:2019}. However, DCS in the visible and UV across comparable bandwidths to conventional spectrometers requires (i) the challenging nonlinear generation of frequency combs across multiple harmonic orders \cite{Lesko:2022,Di:2023,Wu:2023} as well as (ii) methods for detecting and processing DCS signals over a wide spectral range. Due to inadequate optical powers and challenging phase control, UV DCS itself remains largely unexplored, with only a few recent examples \cite{Xu:2023,Bernhardt:2023,Zhang:2023}. 

This work addresses these challenges and achieves multi-harmonic DCS in the UV, visible, and near-infrared (NIR) simultaneously, filling a crucial gap in broad bandwidth and high-resolution (100~MHz) UV-visible spectroscopy. We employ intense few-cycle pulses from erbium fiber lasers (Er:fiber) to generate NIR to UV frequency combs from 380-800~nm in a single bulk crystal. The high peak intensities enable the cascaded generation of several bright harmonics, as shown in Fig. \ref{fig:1}. A spectral multiplexing DCS setup yields simultaneous multi-octave detection over these harmonics with high signal-to-noise and comb mode resolution. The same few-cycle Er:fiber setup has been previously used for high-harmonic generation (HHG) in solids to produce frequency combs spanning the entire UV-visible down to 200~nm, illustrating a direct path to DCS with broader bandwidth coverage at shorter wavelengths. In linking few-cycle harmonic generation to multi-wavelength DCS with over 160,000 resolved comb modes at 100~MHz, our work establishes a foundation for spectroscopy with unparalleled resolution over the full UV-visible range to enable the complete mapping of electronic bands in numerous species.

\begin{figure*}[ht!]
\centering\includegraphics[width=\textwidth]{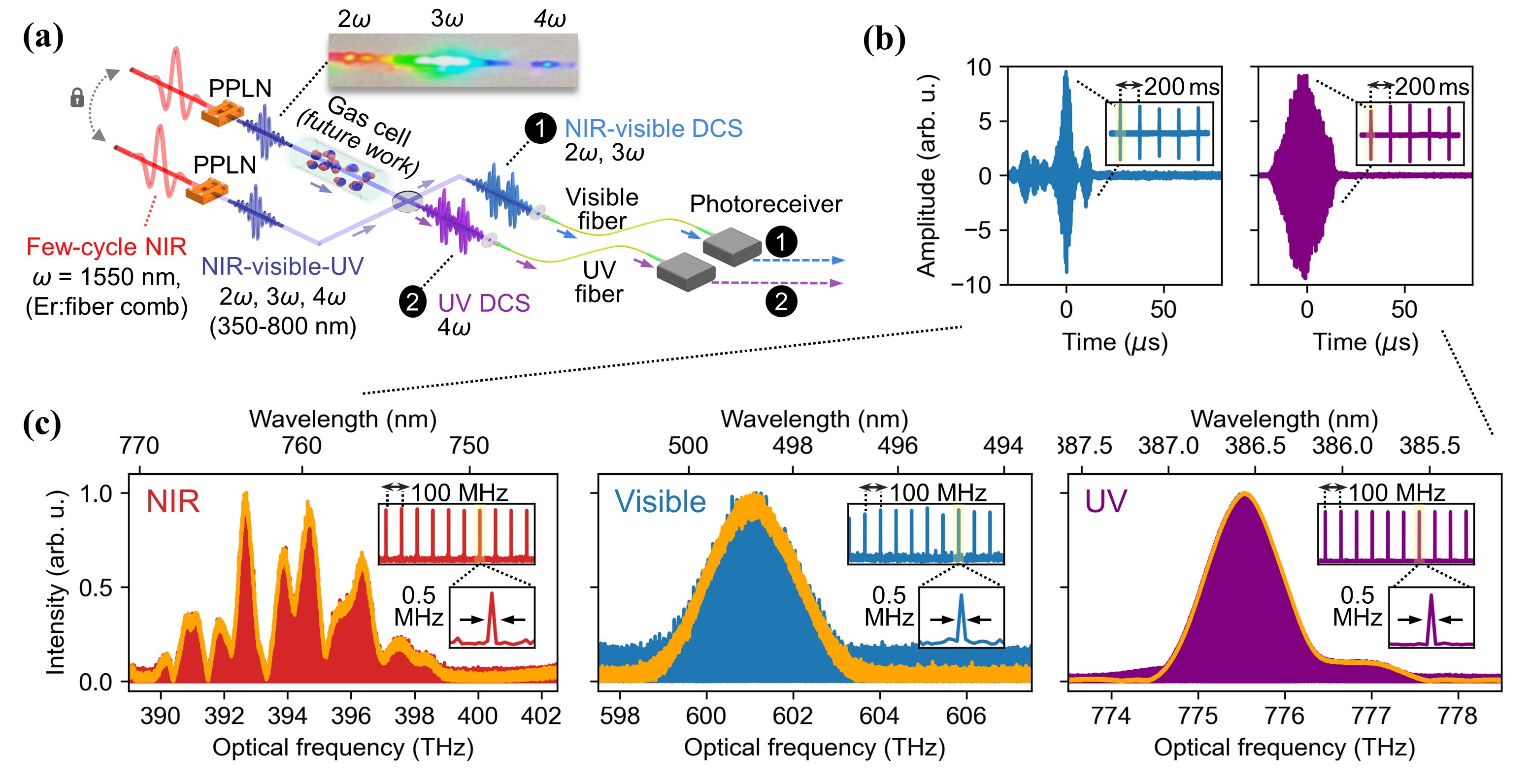}
\caption{NIR-UV dual comb spectrometer and results. (a) Multi-harmonic spectrometer setup. (b) Interferogram streams obtained from visible-NIR (blue curve) and UV (purple curve) dual-comb channels. (c) Comb mode-resolved spectra in the NIR (red plot), visible (blue plot) and UV (purple plot) obtained from 20~s interferogram streams. The orange traces are dual-comb spectra resulting from a coherent average of 500 self-corrected interferograms over a 100~s measurement time.}
\label{fig:2}
\end{figure*}

\begin{figure}[ht!]
\centering
\includegraphics[width=3in]{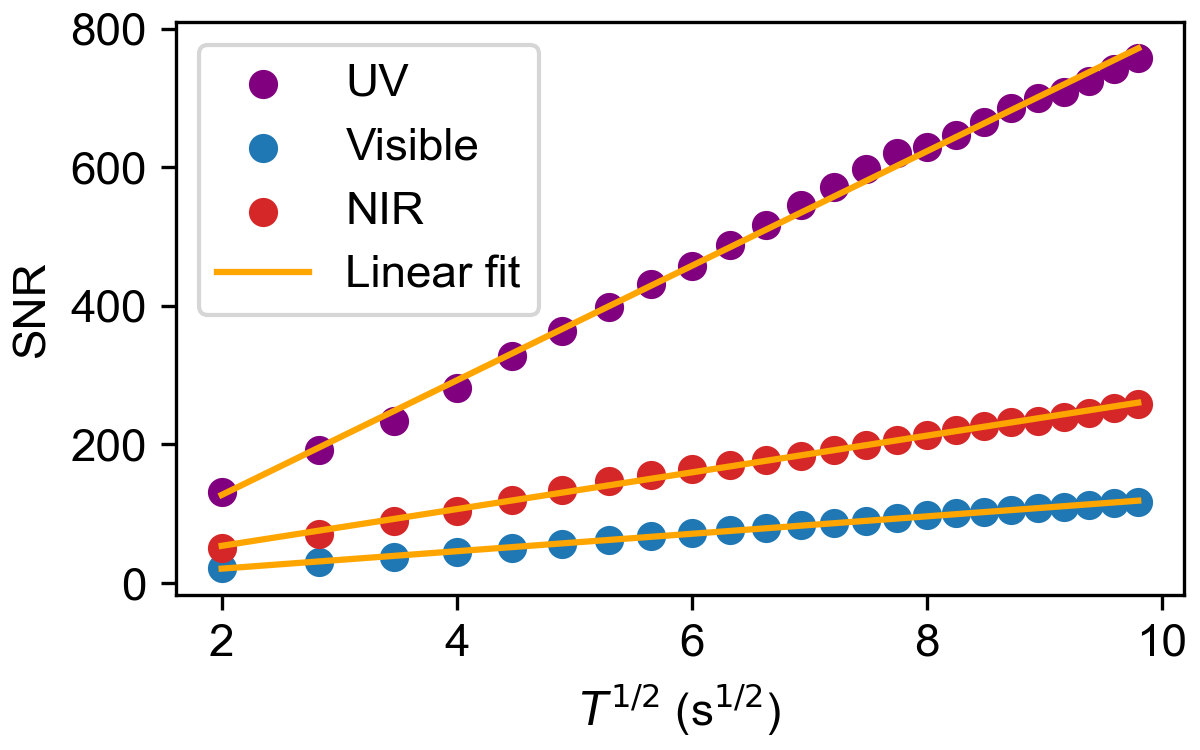}
\caption{Frequency domain signal-to-noise ratios ($SNR$) as a function of the square root of the measurement time ($T^{1/2}$).}
\label{fig:3}
\end{figure}

\begin{figure}[ht!]
\centering
\includegraphics[width=3in]{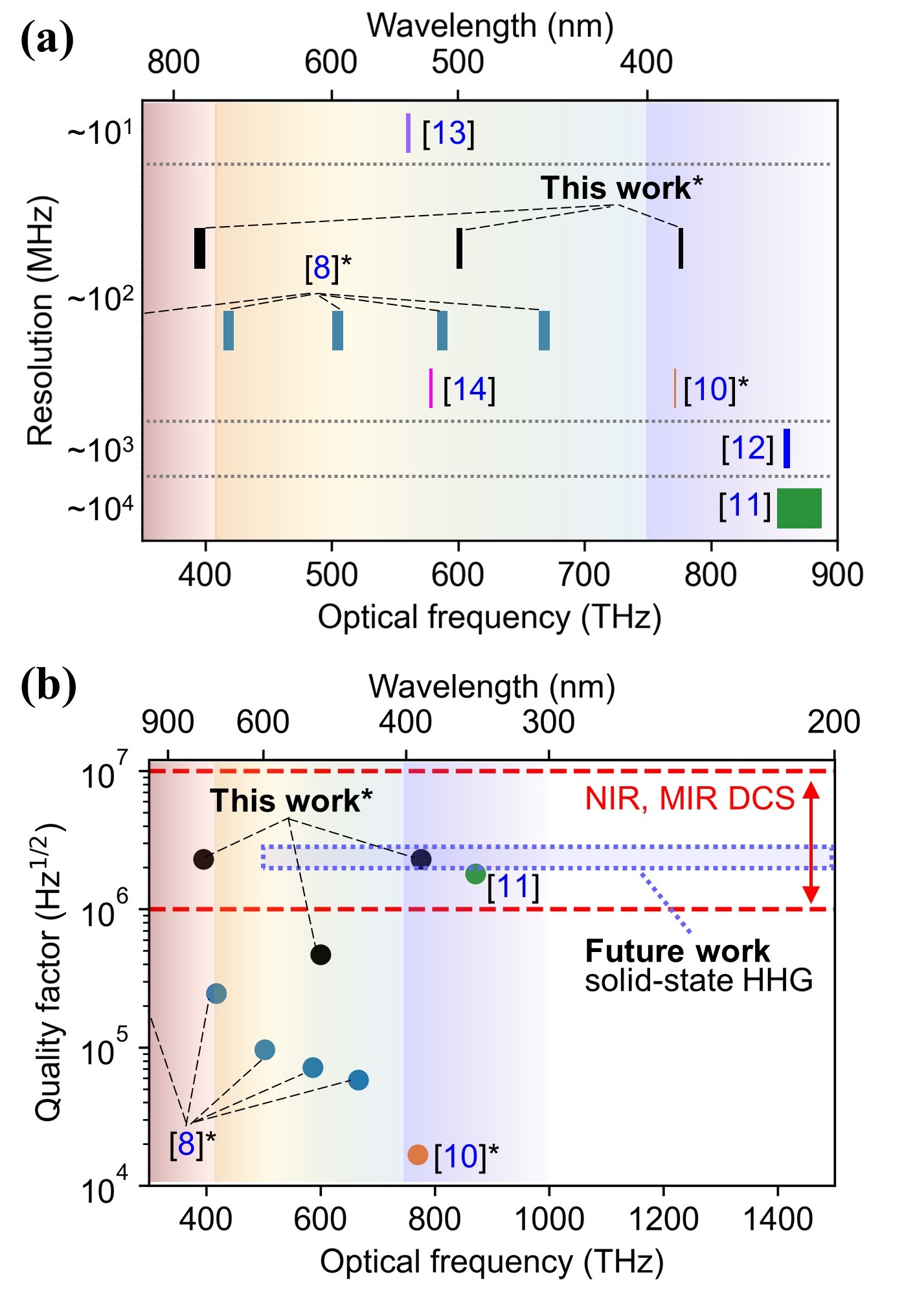}
\caption{Literature comparisons: (a) Overview of resolution and (b) quality factors for UV and visible dual-comb studies \cite{Di:2023, Xu:2023, Bernhardt:2023, Zhang:2023, Sugiyama:2023, Ideguchi:2012}. UV, visible, and NIR regions are shaded in red, orange-blue, and purple, respectively. Typical MIR and NIR dual-comb quality factor ranges are denoted in red. A quality factor estimate for future dual-comb from 200-600 nm is shown in a dashed purple box. * Indicates comb-mode resolved studies.}
\label{fig:4}
\end{figure}

The DCS setup shown in Fig. \ref{fig:2}(a) is based on a pair of repetition rate and phase-stabilized Er:fiber frequency combs at 1550~nm (193~THz) and $\sim$100~MHz repetition rate. We employ the amplification and compression procedure of Ref. \cite{Lesko:2021} to generate near transform-limited 20 fs pulses at $\sim$10~nJ/pulse. NIR-UV combs are generated by focusing the few-cycle pulses in a magnesium oxide-doped periodically-poled lithium niobate (PPLN) crystal (L = 0.5 mm, $\Lambda$ = 20.9 $\mu$m) with 0.9 TW/cm\textsuperscript{2} peak intensities. Supplement \ref{sec:supp} includes simulations confirming that $\chi^{(2)}$ processes dominate the mechanism of harmonic generation through Type-0 phase matching. Cascaded $\chi^{(2)}$ generation produces 2$\omega$ (760~nm, 395~THz), 3$\omega$ (500~nm, 601~THz), and 4$\omega$ (386~nm, 776~THz) harmonics in discrete regions of the NIR, visible, and UV, respectively. The harmonic spectrum is plotted in Fig. \ref{fig:1}, and an image of the spectrally dispersed harmonics is shown in Fig. \ref{fig:2}(a). 

Improvements in spectral overlap between the pair of combs lead to favorable increases in the signal-to-noise ratio ($SNR$) of DCS. In the optical setup, the center frequency and structure of each harmonic can be tuned by adjusting the position of the PPLN in the focus of the beam and the spectrum of the few-cycle pulse. Ultimately, experimental parameters are tuned to maximize the spectral overlap of the 4$\omega$ UV harmonics of each comb at the sacrifice of optimal spectral overlap at the 2$\omega$ NIR and 3$\omega$ visible harmonics. A comparison of the two comb spectra is shown in Supplement \ref{sec:supp}.

As shown in Fig. \ref{fig:2}(a), NIR-UV DCS is performed by heterodyning two combs that are combined using a 50/50 beamsplitter. DCS detection employs a wavelength-multiplexed setup with distinct channels for different wavelength ranges. The beamsplitter generates parallel channels focused into visible and UV single-mode fibers, creating dedicated NIR-visible and UV DCS channels. Subsequently, streams of NIR-visible and UV DCS interferograms are simultaneously measured through fast silicon photodetectors and digitized at 100~MHz using a high-speed 16-bit analog-to-digital card.

The combination of single-mode fibers and the wavelength-multiplexed detection scheme provides several advantages. Firstly, the use of single-mode fiber enhances DCS $SNR$ by ensuring high-quality spatial mode overlap between heterodyning combs. Secondly, exploiting the chromatic focal shift from the aspheric lens enables flexible tuning of frequency coupling efficiency by adjusting the distance from the lens to the fiber entrance. This selective frequency coupling facilitates the removal of the residual fundamental laser $\omega$ and ensures specific coupling of 2$\omega$-3$\omega$ into the NIR-visible DCS channel and 4$\omega$ into the UV DCS channel. Additionally, setting the distance between the aspheric lens and the entrance of the visible single-mode fiber allows for equal transmission powers of 2$\omega$ and 3$\omega$, enabling the adjustment of relative powers of the 2$\omega$ and 3$\omega$ harmonics to prevent detector saturation. Finally, the convenient multiplexing scheme enabled by single-mode fiber coupling allows for parallel acquisition on separate detectors. Since the experimental $SNR$ is limited by relative-intensity noise (RIN), the use of multiple detectors increases the $SNR$ by reducing the bandwidth on each detector \cite{Newbury:2010,Coddington:2016}.

Fig. \ref{fig:2}(b) shows a continuous stream of recorded interferograms from NIR-visible and UV DCS. To prevent aliasing in both channels, a small repetition rate difference $\Delta f_{r}$ of 5~Hz is used, corresponding to sampling in the first Nyquist window and resulting in an interferogram spacing of 1/$\Delta f_{r}$ = 200~ms in the time domain. Continuous streams of interferograms are recorded for 100~s (500 interferograms) and subsequently post-processed to correct for phase and repetition rate drifts using a self-correction algorithm \cite{Hebert:2017,Yu:2019}. Finally, we note that UV DCS interferograms were also collected at $\Delta f_{r}$ = 171~Hz, the largest repetition rate difference that could be adopted without aliasing, but resulted in no substantial difference in $SNR$ for the same measurement time.

Spectra demonstrating NIR-UV DCS at the 2$\omega$ to 4$\omega$ harmonics are plotted in Fig. \ref{fig:2}(c). The spectra result from a Fourier transform of 500 coherently-averaged interferograms and a 20~s, 100 interferogram stream. Zooming into the center of the DCS spectra from 20 s streams reveals individual comb beatnotes spaced by the 100~MHz laser repetition rate. The beatnotes are transform-limited (<0.5~MHz optical width, <25~mHz radio-frequency width), illustrating the combined effectiveness of comb stabilization and post-processing for comb mode resolution. In addition, this demonstration confirms that the mutual coherence of the combs is maintained after multiple nonlinear processes including few-cycle pulse generation and cascaded harmonic generation at high peak intensities.

Table \ref{tab:1} summarizes dual-comb bandwidths and frequency domain $SNR$ for a 100 s measurement. The reduction in DCS bandwidth at higher frequencies aligns with the decreased bandwidth observed for higher multiplicity harmonic orders, as illustrated in Fig. \ref{fig:1}. Notably, the $SNR$ in the UV surpasses that in the NIR-visible. The disparity is attributed to the broader bandwidth of the NIR-visible channel measured on a single detector \cite{Newbury:2010}, reducing its $SNR$. Additionally, a spectral shape mismatch between the two combs in the NIR-visible contributes to diminished $SNR$ in this region (see Supplement \ref{sec:supp}). In the future, the $SNR$ in the NIR and visible can be enhanced by measuring NIR and visible DCS on separate detectors. Nevertheless, $SNR$ levels exceeding 100 for a 100 s measurement are comparable or greater than those recently achieved by UV and visible DCS \cite{Di:2023,Bernhardt:2023,Xu:2023}, and suggests that the spectrometer is suited for simultaneous NIR-UV DCS on gas samples.

\begin{table}[htbp]
\centering
\caption{Dual-comb spectrometer performance, with bandwidth values derived from the full spectral range in each wavelength region. $SNR$ values represent frequency domain signal-to-noise ratios for a 100~s measurement.}
\begin{tabular}{cccccc}
\hline
Harmonic & Bandwidth & SNR & Quality Factor \\
\hline
2$\omega$ (NIR) & 9~THz & 259 & $2.3x10^6 \sqrt{\text{Hz}}$ \\
3$\omega$ (visible) & 4~THz & 117 & $4.7x10^5 \sqrt{\text{Hz}}$\\
4$\omega$ (UV) & 3~THz & 757 & $2.3x10^6 \sqrt{\text{Hz}}$  \\
\hline
\end{tabular}
\label{tab:1}
\end{table}

In Fig. \ref{fig:3}, we illustrate the frequency domain $SNR$ scaling with the square root of the measurement time $T^{1/2}$ up to $T = 100$~s. Linear fits in each spectral regions yield $R^2$ > 0.99. These results confirm the linear scaling of $SNR$ with $T^{1/2}$, affirming the preservation of mutual coherence between the two combs over $T = 100$~s \cite{Newbury:2010,Coddington:2016}. The continued linearity of $SNR$ scaling up to 100~s also implies that longer measurement times could enable even higher $SNR$, at the expense of higher data storage and greater computational time for post-processing.

Fig. \ref{fig:4} provides an overview of all known visible and UV DCS demonstrations to date. In Fig. \ref{fig:4}(a), we compare the spectral resolutions and frequency ranges achieved in previous studies with those in this work. Six studies, including three in the UV \cite{Xu:2023, Bernhardt:2023, Zhang:2023} and three in the visible \cite{Ideguchi:2012, Di:2023}, are considered. Among these, only one conducted DCS over a multi-wavelength range, though not simultaneously \cite{Di:2023}. Specifically, Ref. \cite{Di:2023} demonstrated 108.4~MHz comb mode-resolved DCS in the mid-infrared (MIR), NIR, and visible using discrete harmonics from 450-3750~nm generated in a PPLN waveguide with a MIR pump. While a UV harmonic at 395~nm was generated, its power was insufficient for spectroscopy. In contrast, this work showcases DCS at shorter wavelengths by leveraging the generation of optically bright UV harmonics, enabling simultaneous access to the UV and visible with 100~MHz resolution. Notably, among the three previous UV DCS demonstrations \cite{Xu:2023, Bernhardt:2023, Zhang:2023}, one achieved comb mode resolution at 500~MHz over less than 1~THz of bandwidth. Here, we measure 16~THz of simultaneously-resolved bandwidth, corresponding to 160,000 individually resolved comb modes.

Fig. \ref{fig:4}(b) compares the quality factors of DCS examples for which estimates were possible. Quality factors are a metric for comparing the performance of spectrometers, and are defined in analogy to the commonly-used DCS figure of merit \cite{Coddington:2016,Galtier:2020,Bernhardt:2023}: $SNR$ x $M$ / $T^{1/2}$. $SNR$ is the frequency domain signal-to-noise ratio, and $M$ = $\Delta$$\omega$/$\omega_{res}$ where $\Delta\omega$ is the spectral bandwidth, $\omega_{res}$ is the resolution, and $T$ is the measurement time. Values used to calculate quality factors are summarized in Table \ref{tab:1}]. The quality factors fall between $4x10^5$ and $3x10^6$ in each spectral region, approaching those of more established MIR and NIR DCS experiments from $10^6$-$10^7 \sqrt{\text{Hz}}$ \cite{Baumann:2011,Zolot:2012,Coddington:2016,Lind:2020,Liu:2023}. The quality factors of this visible and UV DCS experiments exceed those of recent experiments in the same spectral range \cite{Di:2023,Bernhardt:2023,Xu:2023}.

The few-cycle Er:fiber technology, as demonstrated in Fig.~\ref{fig:1}, provides comprehensive UV-visible spectral coverage at lower powers when applied to high-harmonic generation (HHG) in a solid \cite{Lesko:2022}. Whereas PPLN harmonic generation is limited to $\gtrapprox$330~nm due to the low transparency of PPLN in the UV \cite{Wu:2023}, HHG generates 2$\omega$ through 7$\omega$ harmonics, extending UV coverage to 200~nm, which could enable the deep-UV DCS for the first time. A wavelength-multiplexed scheme with multiple detectors, as employed in this work, would enable simultaneous detection of the harmonics and maximize $SNR$. A quality factor of $2.5x10^6 \sqrt{\text{Hz}}$ over the 200-600~nm span of the comb (see Supplement \ref{sec:supp} for details) is shown in Fig.~\ref{fig:4}(e). This quality factor is equivalent with the results of this work and recent UV and visible DCS studies, indicating the potential of HHG combs for future UV-visible DCS.

In summary, this work demonstrates a DCS spectrometer with simultaneous multi-wavelength coverage from NIR-UV and with high quality factors and $SNR$ ratios. To the best of our knowledge, this is the first DCS demonstration to access both the visible and UV regions. A future extension of the optical setup presented in this work to few-cycle HHG should enable high-resolution, 100~MHz spectroscopy across the entire UV-visible for the first time, representing a key advance in instrumentation for both fundamental and applied spectroscopy.

\vspace{8pt}

\textbf{Funding} U.S. Air Force (FA9550-16-1-0016, FA9550-22-1-0483); National Institute of Standards and Technology (70NANB18H006).

\textbf{Acknowledgments} The authors thank J. Genest, M. Walsh, E. Baumann, R. Cole, I. Coddington, and D. Slichter for valuable discussions and feedback. K.F. Chang acknowledges the National Research Council Research Associate Programs Fellowship for funding. E.J. Tsao acknowledges the National Defense Science and Engineering Graduate Fellowship for funding.

\textbf{Disclosures} The authors declare no conflicts of interest.

\textbf{Data availability} Data in this paper may be obtained from the authors upon reasonable request.

\textbf{Supplemental document}
See Supplement \ref{sec:supp} for supporting content.

\newpage
\section{{\fontsize{20pt}{20pt}\selectfont
    \textbf{Supplement} \label{sec:supp}
}}

\subsection{NIR-UV harmonic generation process}
In this study, a periodically-poled lithium niobate (PPLN) crystal with a fixed poling period of 20.9~$\mu$m and a 0.5 mm thickness was employed for harmonic generation. Experimental results indicate the generation of 2$\omega$ (NIR), 3$\omega$ (visible), and 4$\omega$ (UV) harmonics from linearly-polarized few-cycle pulses oriented for Type 0 second-harmonic generation. A 14\% conversion efficiency across the NIR-UV spectrum is achieved, with a conversion efficiency of 13.8\%, 0.17\%, and 0.01\% to 2$\omega$, 3$\omega$, and 4$\omega$ respectively.

To investigate the harmonic generation mechanism, we analyzed variations in the harmonic spectrum with respect to the average input power of the driving laser. Experimental control over the input power employed a linear polarizer and a half-wave plate before the PPLN crystal. Simulations of harmonic spectra spanning the same input power range were conducted using PyNLO, a GitHub package our group has developed for simulating nonlinear interactions \cite{Fredrick:2023}. The simulations account for experimental parameters including 20 fs input pulses and 0.9 TW/cm\textsuperscript{2} peak intensities, and restricted nonlinear interactions to $\chi(2)$ effects and Type 0 mixing. The PPLN was not temperature-controlled in the experiment, and any heating effects on the crystal were considered negligible for the purposes of the simulation.

\begin{figure}[hbt!]
\centering
\includegraphics[width=3in]{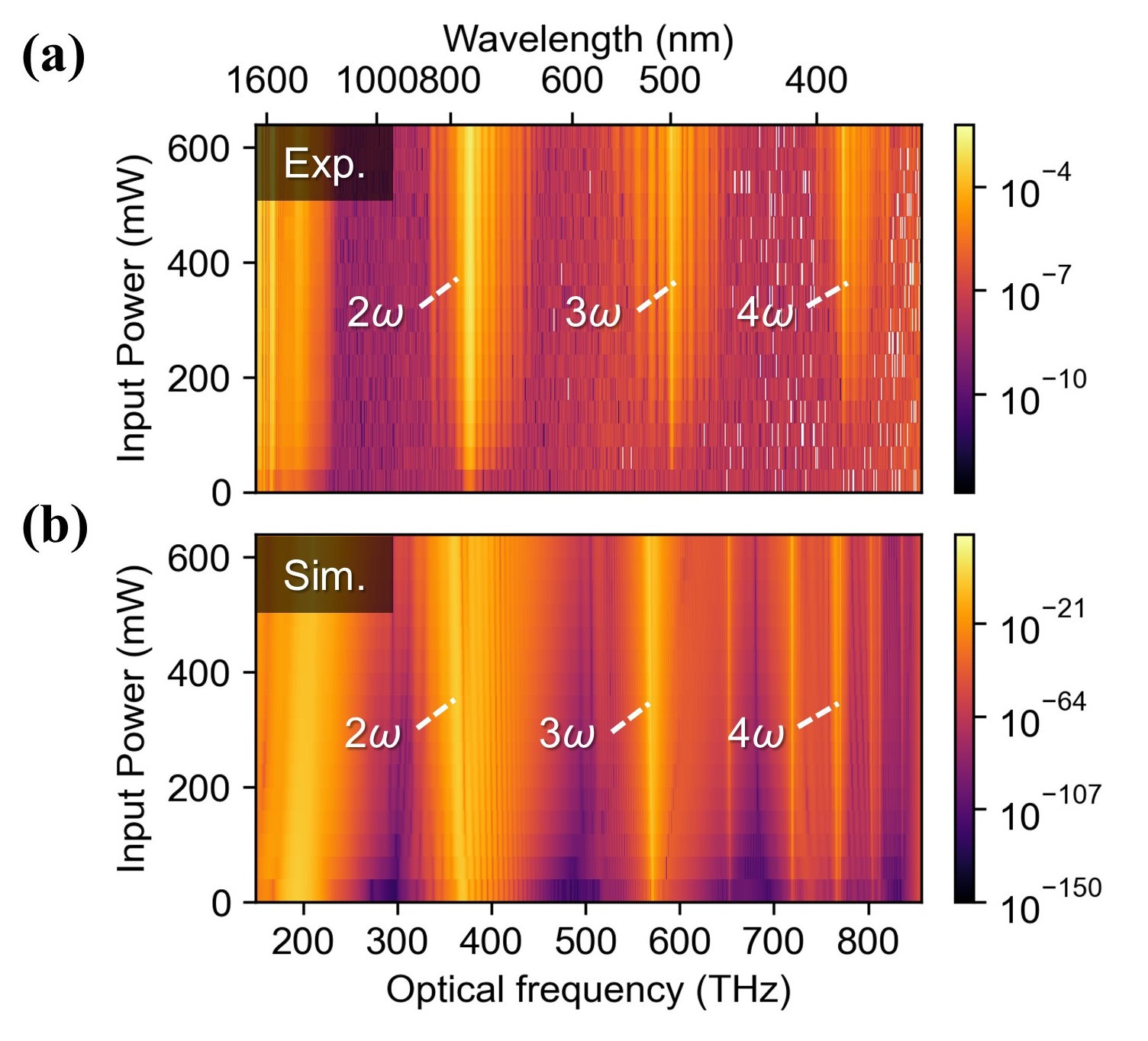}
\caption{NIR-visible harmonic spectra as a function of input power plotted as colormaps. (a) Experimental spectra. (b) Simulated spectra. The colors represent spectral intensity and are plotted on a logarithmic scale in arbitrary units.}
\label{suppfig:1}
\end{figure}

A comparison of experimental and simulated harmonic spectra as a function of input power are plotted in Fig. \ref{suppfig:1}. At a range of input powers, the relative intensities of the harmonics in the experimental spectrum appear to also be captured by the simulation, and suggests that $\chi^{(2)}$ nonlinear optics is sufficient to describe the experimental results. The primary discrepancy between the simulated and experimental results is a red-shift of the simulated 2$\omega$ and 3$\omega$ peaks relative to those of the experiment and the appearance of sharp spectral tones between 600-750 THz and >800 THz in the simulation. This discrepancy may stem from the assumption of the PPLN temperature of $24.5^\circ$ in the simulation, resulting in shifted frequencies.

In both the simulation and experiment, the 3$\omega$ and 4$\omega$ harmonics are weaker and narrower than 2$\omega$. The PPLN poling period quasi-phase matches Type 0 generation of 2$\omega$ from the red edge at 1605 nm in the few-cycle pulse spectrum \cite{Lesko:2021}. However, the generation of 3$\omega$ and 4$\omega$ are not phase-matched by subsequent $\chi(2)$ processes, thereby limiting the bandwidth and conversion efficiency of these harmonics. In the future, a chirped PPLN design could facilitate higher conversion efficiencies and broader bandwidth conversion to the UV-visible.\cite{Chen:2015}. However, the transparency of PPLN is limited to the band gap at approximately 330 nm, and techniques such as high-harmonic generation become necessary to achieve continuous UV-visible coverage out to 200~nm \cite{Lesko:2022}.

\subsection{Comparison of NIR-UV comb spectra}

\begin{figure}[hbt!]
\centering
\includegraphics[width=\textwidth]{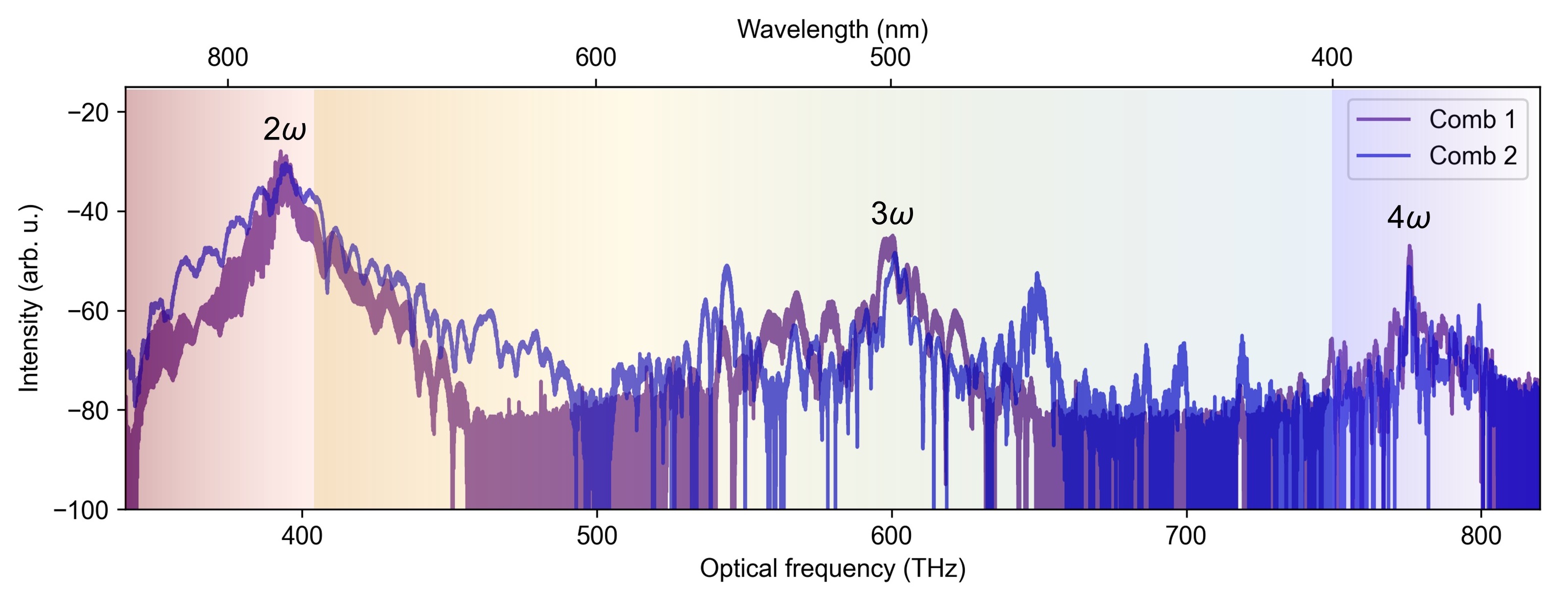}
\caption{Experimental NIR-visible comb spectra. The NIR, visible, and UV are shaded in red, orange-blue, and purple, respectively.}
\label{suppfig:2}
\end{figure}

Fig. \ref{suppfig:2} shows the NIR-UV harmonic spectra of each of the two combs used for the dual-comb spectroscopy (DCS) described in the main text. As shown in the plot, the two combs do not have complete overlap at each harmonic, leading to lower $SNR$ for DCS in spectral regions with reduced overlap, particularly the visible and NIR.

\subsection{Quality factor estimate for a dual-comb spectrometer based on frequency combs from high-harmonic generation}

Fig. 1 of the main text shows a 100 MHz UV-visible frequency comb spectrum generated few-cycle high-harmonic generation (HHG) \cite{Lesko:2022}. The spectrum extends continuously from 200-600 nm (500-1500 THz), and an average quality factor for this wavelength range can be calculated. The formula for calculating quality factors is
\begin{equation}\label{eqn:1}
    Q = \frac{SNR \times M}{T^{1/2}}
\end{equation}
where $SNR$ is the frequency domain signal-to-noise ratio at the peak of the dual-comb signal after the measurement time $T$, and $M$ = $\Delta$$\omega$/$\omega_{res}$ where $\Delta\omega$ is the spectral bandwidth, $\omega_{res}$ is the resolution \cite{Coddington:2016,Di:2023,Bernhardt:2023}.

The $SNR$ value in the quality factor formula is estimated using Eq. 2 of Ref. \cite{Newbury:2010}. In this calculation, a measurement time of $T = 100$~s and a resolution of $\omega_{res} = 100$~MHz is used, and known quantities such as laser power and detector noise-equivalent power are utilized. A wavelength-multiplexed DCS scheme employing three separate photodetectors to enable the detection of two harmonics per detector is assumed, and any mismatched spatial, spectral, and mode overlap between the two combs is considered negligible. Crucially, the $SNR$ calculation utilizes a value of RIN in its estimate. In our previous work on HHG of UV-visible combs \cite{Lesko:2022}, the RIN of the harmonic light was not characterized. However, because the the primary experiment in this work using PPLN and the previous HHG work utilize the same driving lasers at similar peak intensities, we estimate the RIN of the HHG combs from the RIN of the current experiment. Here, we consider the RIN recorded from the photodetection of two harmonics simultaneously, 3$\omega$ and 4$\omega$, which is measured to be $-145$~dBc/Hz. Given this, we conservatively assume $-130$~dBc/Hz for the RIN of HHG combs. Considering these factors, an estimated $SNR$ of 2.5 for a 100~s measurement time is derived for a DCS experiment utilizing HHG combs.

Employing Eq. \ref{eqn:1}, a 100 MHz resolution DCS experiment employing the full bandwidth of the HHG frequency comb can be estimated to have an average quality factor calculated using the following values: $SNR$ = 2.5 for $T$ = 100~s, $\Delta\omega$ = 1000~THz, and $\omega_{res}$ = 100~MHz. Therefore, the calculated quality factor is $Q = 2.5x10^6 \sqrt{\text{Hz}}$.

\bibliography{main}

\begin{thebibliography}{10}
\newcommand{\enquote}[1]{``#1''}

\bibitem{Galtier:2020}
S.~Galtier, P.~Clément, and P.~Rairoux, \enquote{Towards {DCS} in the {UV} spectral range for remote sensing of atmospheric trace gases,} {\protect\JournalTitle{Remote Sensing}} \textbf{12} (2020).

\bibitem{Das:2021}
A.~Das, \emph{Portable UV–Visible Spectroscopy – Instrumentation, Technology, and Applications} (John Wiley \& Sons, Ltd, 2021), chap.~8, pp. 179--207.

\bibitem{Schuster:2021}
V.~Schuster, C.~Liu, R.~Klas, P.~Dominguez, J.~Rothhardt, J.~Limpert, and B.~Bernhardt, \enquote{Ultraviolet dual comb spectroscopy: a roadmap,} {\protect\JournalTitle{Opt. Express}} \textbf{29}, 21859--21875 (2021).

\bibitem{Coddington:2016}
I.~Coddington, N.~Newbury, and W.~Swann, \enquote{Dual-comb spectroscopy,} {\protect\JournalTitle{Optica}} \textbf{3}, 414--426 (2016).

\bibitem{Weichman:2019}
M.~L. Weichman, P.~B. Changala, J.~Ye, Z.~Chen, M.~Yan, and N.~Picqué, \enquote{Broadband molecular spectroscopy with optical frequency combs,} {\protect\JournalTitle{Journal of Molecular Spectroscopy}} \textbf{355}, 66--78 (2019).

\bibitem{Picque:2019}
N.~Picque and T.~Hansch, \enquote{Frequency comb spectroscopy,} {\protect\JournalTitle{Nat. Photon.}} \textbf{13}, 146--–157 (2019).

\bibitem{Lesko:2022}
D.~Lesko, K.~Chang, and S.~Diddams, \enquote{High-sensitivity frequency comb carrier-envelope-phase metrology in solid state high harmonic generation,} {\protect\JournalTitle{Optica}} \textbf{9}, 1156--1162 (2022).

\bibitem{Di:2023}
Y.~Di, Z.~Zuo, D.~Peng, D.~Luo, C.~Gu, and W.~Li, \enquote{Dual-comb spectroscopy from the ultraviolet to mid-infrared region based on high-order harmonic generation,} {\protect\JournalTitle{Photon. Res.}} \textbf{11}, 1373--1381 (2023).

\bibitem{Wu:2023}
T.-H. Wu, L.~Ledezma, C.~Fredrick, P.~Sekhar, R.~Sekine, Q.~Guo, R.~M. Briggs, A.~Marandi, and S.~A. Diddams, \enquote{Visible to ultraviolet frequency comb generation in lithium niobate nanophotonic waveguides,} {\protect\JournalTitle{arXiv}} \textbf{2305.08006} (2023).

\bibitem{Xu:2023}
B.~Xu, Z.~Chen, T.~W. Hänsch, and N.~Picqué, \enquote{Near-ultraviolet photon-counting dual-comb spectroscopy,} {\protect\JournalTitle{arXiv}} \textbf{2307.12869} (2023).

\bibitem{Bernhardt:2023}
B.~Bernhardt, L.~Furst, A.~Kirchner, A.~Eber, F.~Siegrist, and R.~di~Vora, \enquote{Broadband near-ultraviolet dual comb spectroscopy,} {\protect\JournalTitle{Research Square}} \textbf{10.21203/rs.3.rs-2760097/v1} (2023).

\bibitem{Zhang:2023}
Y.~Zhang, J.~J. McCauley, and R.~J. Jones, \enquote{Ultraviolet dual-comb spectroscopy utilizing intra-cavity high harmonic generation,} {\protect\JournalTitle{CLEO 2023}} \textbf{SF3F.6} (2023).

\bibitem{Sugiyama:2023}
Y.~Sugiyama, T.~Kashimura, K.~Kashimoto, D.~Akamatsu, and F.-L. Hong, \enquote{Precision dual-comb spectroscopy using wavelength-converted frequency combs with low repetition rates,} {\protect\JournalTitle{Scientific Reports}} \textbf{13}, 2549 (2023).

\bibitem{Ideguchi:2012}
T.~Ideguchi, A.~Poisson, G.~Guelachvili, T.~W. H\"{a}nsch, and N.~Picqu\'{e}, \enquote{Adaptive dual-comb spectroscopy in the green region,} {\protect\JournalTitle{Opt. Lett.}} \textbf{37}, 4847--4849 (2012).

\bibitem{Lesko:2021}
D.~Lesko, H.~Timmers, S.~Xing, A.~Kowligy, A.~Lind, and S.~Diddams, \enquote{A six-octave optical frequency comb from a scalable few-cycle erbium fibre laser,} {\protect\JournalTitle{Nat. Photon.}} \textbf{15}, 281–286 (2021).

\bibitem{Newbury:2010}
N.~R. Newbury, I.~Coddington, and W.~Swann, \enquote{Sensitivity of coherent dual-comb spectroscopy,} {\protect\JournalTitle{Opt. Express}} \textbf{18}, 7929--7945 (2010).

\bibitem{Hebert:2017}
N.~B. H\'{e}bert, J.~Genest, J.-D. Desch\^{e}nes, H.~Bergeron, G.~Y. Chen, C.~Khurmi, and D.~G. Lancaster, \enquote{Self-corrected chip-based dual-comb spectrometer,} {\protect\JournalTitle{Opt. Express}} \textbf{25}, 8168--8179 (2017).

\bibitem{Yu:2019}
H.~Yu, K.~Ni, Q.~Zhou, X.~Li, X.~Wang, and G.~Wu, \enquote{Digital error correction of dual-comb interferometer without external optical referencing information,} {\protect\JournalTitle{Opt. Express}} \textbf{27}, 29425--29438 (2019).

\bibitem{Baumann:2011}
E.~Baumann, F.~R. Giorgetta, W.~C. Swann, A.~M. Zolot, I.~Coddington, and N.~R. Newbury, \enquote{Spectroscopy of the methane ${\ensuremath{\nu}}_{3}$ band with an accurate midinfrared coherent dual-comb spectrometer,} {\protect\JournalTitle{Phys. Rev. A}} \textbf{84}, 062513 (2011).

\bibitem{Zolot:2012}
A.~M. Zolot, F.~R. Giorgetta, E.~Baumann, J.~W. Nicholson, W.~C. Swann, I.~Coddington, and N.~R. Newbury, \enquote{Direct-comb molecular spectroscopy with accurate, resolved comb teeth over 43 thz,} {\protect\JournalTitle{Opt. Lett.}} \textbf{37}, 638--640 (2012).

\bibitem{Lind:2020}
A.~J. Lind, A.~Kowligy, H.~Timmers, F.~C. Cruz, N.~Nader, M.~C. Silfies, T.~K. Allison, and S.~A. Diddams, \enquote{Mid-infrared frequency comb generation and spectroscopy with few-cycle pulses and ${\ensuremath{\chi}}^{(2)}$ nonlinear optics,} {\protect\JournalTitle{Phys. Rev. Lett.}} \textbf{124}, 133904 (2020).

\bibitem{Liu:2023}
M.~Liu, R.~M. Gray, L.~Costa, C.~R. Markus, A.~Roy, and A.~Marandi, \enquote{Mid-infrared cross-comb spectroscopy,} {\protect\JournalTitle{Nature Communications}} \textbf{14}, 1044 (2023).

\bibitem{Fredrick:2023}
C.~Fredrick, \enquote{{PyNLO},} \url{https://github.com/cdfredrick/PyNLO} (2023).

\bibitem{Chen:2015}
B.-Q. Chen, C.~Zhang, C.-Y. Hu, R.-J. Liu, and Z.-Y. Li, \enquote{High-efficiency broadband high-harmonic generation from a single quasi-phase-matching nonlinear crystal,} {\protect\JournalTitle{Phys. Rev. Lett.}} \textbf{115}, 083902 (2015).

\end{thebibliography}

\end{document}